    \documentstyle[12pt]{article}          

\def\Imag{\mathop{\rm Im}\nolimits}

\makeatletter 
  \@ifundefined{th@half}{%
    \gdef\preprintno#1{\gdef\@preprintno{#1}}\gdef\@preprintno{}
    \gdef\abstracttext#1{\gdef\@abstract{#1}}\gdef\@abstract{}}{}
  \@ifundefined{landscape}{}{\landscape}
\makeatother

\title{%
  Unimagined Imaginary Parts in Heavy Quark Effective Field Theory}

\author{%
  Wolfgang Kilian%
    \thanks{e-mail: {\tt kilian@crunch.ikp.physik.th-darmstadt.de}}
  {\small and} Thomas Mannel%
    \thanks{e-mail: {\tt mannel@crunch.ikp.physik.th-darmstadt.de}}\\
    Technische Hochschule Darmstadt \\
    D-6100 Darmstadt \\
    Germany \\
  \hfill\\
  Thorsten Ohl%
    \thanks{e-mail: {\tt ohl@physics.harvard.edu}}\\
    Lyman Laboratory of Physics\\
    Harvard University \\
    Cambridge, MA 02138}

\preprintno{\#HUTP-93/A009\\IKDA 93/9}
\date{3/93}

\abstracttext{%
   We argue that the imaginary parts of the anomalous dimensions in
   the multiparticle sectors of heavy quark effective field theory may
   be removed by a suitable redefinition of the multiparticle states.
   The connection between the imaginary parts of the anomalous
   dimensions and the interquark potential is pointed out.}

\begin{document}
\maketitle

\makeatletter 
  \@ifundefined{th@half}{\begin{abstract}\@abstract\end{abstract}}{}
  \@ifundefined{starttext}{}{\starttext}
\makeatother

\section{Introduction}
\label{sec:intro}

Heavy quark effective field theory
(HqEFT)~\cite{IW88,IW89,EH90,Gri90,Geo90} has proven to
be a useful tool for the systematic analysis of systems containing
heavy quarks. This approach is based on the heavy mass
limit in which the Hilbert space decomposes into superselection
sectors labelled by the velocities of the heavy quarks.
Corrections to the leading behavior may be calculated systematically
and we shall focus in this work on the strong interaction corrections.
These are treated in renormalization group improved perturbation
theory, resumming logarithms of the heavy quark masses.
Weak interactions will introduce operators connecting different
velocity sectors and in general their anomalous dimensions will depend
on the velocities of the heavy quarks. The lowest order expressions
have been given in~\cite{FGGW90,GKMW91}.

In the limit~$m_Q \to \infty$ particle and antiparticle number are
separately conserved and most of the applications discussed up to now
deal with the one (anti)particle sector of HqEFT. However, in the two
particle sector some of the anomalous dimensions develop imaginary
parts. In the full theory imaginary parts of Green's functions emerge
{}from analytic continuation and some of these imaginary parts contain
logarithms of the masses. In the effective theory these logarithms are
reproduced by the renormalization group.  Therefore the anomalous
dimensions have to pick up imaginary parts because of analyticity
of the full theory.

Complex anomalous dimensions are somewhat unusual and the purpose of
the present letter is to shed some light on their origin.  We shall
argue that the imaginary parts may be removed by a suitable
redefinition of the multiparticle states of HqEFT. We shall construct
these redefined  multiparticle states to order~$\alpha_s$ explicitly
in the next section and discuss the extension of this method to all orders.

In the two particle state the imaginary parts of the anomalous
dimensions lead to phases of the Wilson coefficients. Upon closer
inspection, these phases turn out to be the nonabelian generalizations
of the Coulomb phases familiar from QED~\cite{Dol64,KF70}.
In a similar way
as it may be done for the QED case, one may derive an interquark
potential suitable for the physics at the renormalization scale. This
is done up to two loops in section~\ref{sec:pot}.

\section{Complex Anomalous Dimensions}
\label{sec:anodim}

Since HqEFT conserves particle and antiparticle
number separately, we shall begin our discussion in the
particle-particle and particle-antiparticle sectors.
An effective Hamiltonian
\begin{equation}
  H_{{\rm eff}} = \sum_{i=1}^n \eta_i (\mu) {\cal O}_i (\mu)
\end{equation}
having matrix elements with two heavy
particles in the final state will be expanded in a basis of hermitian
operators~${\cal O}_i, i = 1,\dots,n$ with Wilson coefficients~$\eta_i$.
They are calculated from matching at the heavy quark mass scale and
subsequent renormalization group running.
In such a basis the anomalous dimensions
will be given by a specific~$n\times n$ matrix~$\gamma$.  The
hermiticity of the effective interaction requires real entries for
the matrix~$\gamma$.  However, in HqEFT imaginary parts appear as
coefficients of logarithmic divergences as soon as the corresponding
operator has matrix elements with a two- or more heavy particle
state~\cite{GKMW91}. This hints at a possible problem for the
interpretation of HqEFT for these final states.
The imaginary parts of these anomalous dimensions
have been calculated to leading
order in~\cite{GKMW91}.  The generic expression is given by
\begin{equation}
\label{eq:imag}
  \Imag \gamma(\alpha,vv') = - \alpha \frac{vv'}{\sqrt{(vv')^2-1}}
     T^a\otimes T^a.
\end{equation}
The color matrix~$T^a\otimes T^a$ acts on the two heavy quarks in the
final state. Coupling the two final state heavy quarks to definite
color, it will be sufficient to consider the following two sets of
operators.
In the particle-antiparticle sector we have a color singlet and
a color octet operator
\begin{eqnarray}
  J^1 & = & (\bar h_\alpha(v) \psi) (\bar\phi h_\alpha(v')) \\
  J^{8,a} & = & (\bar h_\alpha(v) \psi) T^a_{\alpha\beta}
     (\bar\phi h_\beta(v'))
\end{eqnarray}
and similarly in the particle-particle sector there is a color
antitriplet and a color sextet
\begin{eqnarray}
  J^{\bar 3}_{\alpha\beta} & = & (\bar\psi h_\alpha(v))
          (\bar\phi h_\beta(v'))
     - (\bar\psi h_\alpha(v)) (\bar\phi h_\beta(v')) \\
  J^6_{\alpha\beta} & = & (\bar\psi h_\alpha(v)) (\bar\phi h_\beta(v'))
     + (\bar\psi h_\alpha(v)) (\bar\phi h_\beta(v')) .
\end{eqnarray}
In these equations,~$\alpha,\beta = 1,\ldots,3$
and~$a = 1,\ldots,8$ are the color indices. The Dirac spinors~$\psi$
and~$\phi$ are determined by the specific form of the
operators~${\cal O}_i$; they may contain fields corresponding to
the light degrees of freedom and possibly
another heavy quark in the initial state.

In this basis
the color matrix~$T^a\otimes T^a$ is diagonal which means that
the imaginary part of the anomalous dimension is diagonal and
may be written as
\begin{displaymath}
  \Imag \gamma(\alpha,vv') = \alpha \frac{vv'}{\sqrt{(vv')^2-1}} K_C,
\end{displaymath}
where the color factor~$K_C$ is determined by the action
of~$T^a\otimes T^a$ on the operators~$J_{1,8,\bar3,6}$, yielding
the eigenvalues~$4/3$, $-1/6$, $2/3$, and~$-1/3$.

Thus the imaginary parts in fact depend only on the total color
of the heavy quarks in the final state. This suggests that a suitable
redefinition of the final state may render the
anomalous dimensions real.

\section{Redefinition of the Multiparticle States}
\label{sec:redef}

Before we consider the nonabelian case, we shall briefly review the
exactly soluble abelian case which has been extensively studied
in the context of the QED infrared problem~\cite{BN37,KF70}.
The interaction in the
abelian case is given by
\begin{equation}
  H_I (x_0) = g \int d^3 \vec{x} j_\mu (x) A^\mu (x),
\end{equation}
where the current of a fermion~$Q$ is given by
\begin{equation} \label{eq:jmu}
j_\mu (x) = \bar{Q} (x) \gamma_\mu Q (x) .
\end{equation}
In the nonrecoil limit corresponding to the heavy mass limit
for~$Q$ the current~(\ref{eq:jmu}) may be rewritten~\cite{KF70}
\begin{equation} \label{eq:Jmu}
j_\mu (x) \to J_\mu (x) =  \int\!d^3 \vec{v}\, v_\mu\,
          \delta^3 (\vec{x} - \vec v x_0 / v_0) \, n(v),
\end{equation}
where~$v = p_Q / m_Q$ is the velocity of the heavy particle and
\begin{equation}
n(v) =  \sum_{s=\pm} \left( b^\dagger (v,s) b (v,s) -
                     d^\dagger (v,s) d (v,s) \right)
\end{equation}
is the charge density operator with~$b(v,s),d(v,s)$
being the annihilation operators for particles and antiparticles
of spin~$s$
moving with velocity~$v$ respectively. The dynamics of the system
governed by the Hamiltonian
\begin{equation}
{\cal H}_I (x_0) = g \int d^3 \vec{x} J_\mu (x) A^\mu (x)
\end{equation}
is exactly soluble. The operator~$U$ transforming a free state into
an interacting one is given by~\cite{KF70}
\begin{equation} \label{eq:U}
  U = \exp (iR) \exp (i\Omega),
\end{equation}
with the radiation operator
\begin{equation}
  R = - g \int d^4 x \Theta (-x_0) J_\mu (x) A^\mu (x)
\end{equation}
and phase operator
\begin{equation}
  \Omega  =  \frac{g^2}{2} \int d^4 x d^4 y
      \Theta(-x_0) \Theta(x_0-y_0)
         \left[J_\mu (x) A^\mu (x)\,,\, J_\nu (y) A^\nu (y) \right],
\end{equation}
This model is exactly soluble because in the abelian case the
current~(\ref{eq:Jmu}) commutes with itself for all~$x$ and~$y$
$$
\left[J_\mu (x) \,,\, J_\nu (y)  \right] = 0 .
$$
Thus the phase operator may also be written as
\begin{equation}
  \Omega  =  \frac{g^2}{2} \int d^4 x d^4 y
      \Theta(-x_0) \Theta(x_0-y_0)
         J_\mu (x) \left[A^\mu (x)\,,\, A^\nu (y) \right] J_\nu (y).
\end{equation}
Normal ordering of the two currents introduces a mass renormalization
term which is dealt with as usual. Inserting the~$c$-number commutator
of the~$A$ fields yields for the phase operator
\begin{equation}
\Omega  =  - \frac{g^2}{8 \pi} \int d^3 \vec{v} d^3 \vec{v}\,'
           \frac{vv'}{\sqrt{(vv')^2-1}} {:}n(v)n(v'){:}
           \int \frac{dx_0}{x_0}.
\end{equation}
The divergent~$x_0$ integration is cut off at small~$x_0$ by the
inverse of the mass~$m_Q$ of the heavy particle and at large~$x_0$
by a scale~$\mu$ which will turn out to be the renormalization
scale~$\mu$. Thus we have
\begin{equation}
\Omega  =  - \frac{\alpha}{2} \int d^3 \vec{v} d^3 \vec{v}\,'
           \frac{vv'}{\sqrt{(vv')^2-1}} {:}n(v)n(v'){:}
           \ln \left(\frac{m_Q}{\mu} \right) .
\end{equation}
In the abelian case this operator will contribute to~$U$ as a phase
factor multiplying the multiparticle states. The transformation
$$
U_\Omega = \exp (i\Omega)
$$
is unitary by itself and yields the correct Coulomb phases familiar
{}from QED if applied to a multiparticle state~\cite{KF70}.

On the other hand, one may also calculate the anomalous dimensions
in the abelian version of HqEFT with no light quarks which exactly
corresponds to the above model. From the exact solution~(\ref{eq:U})
we read off that the anomalous dimensions are given by the
one loop result. The~$\beta$ function is trivial ($\beta\equiv0$) since
there is no self coupling of the gauge field and no light particles.
However, the anomalous dimensions in this simple model develop an
imaginary part similar to the one in HqEFT which we now trace back
to the operator~$U_\Omega$.

In fact, the imaginary part of the anomalous dimensions yields a
phase factor in the Wilson coefficient.
This phase factor for a matrix element~${\cal M}(vv')$ with
two heavy particles moving with velocities~$v$ and~$v'$ respectively
in the final state is governed by a
renormalization group equation with only the imaginary part
of the anomalous dimension~\cite{GKMW91}
\begin{equation} \label{eq:RGe}
\left( \mu \frac{\partial}{\partial \mu} + i \Imag \gamma \right)
{\cal M} (vv') = 0 .
\end{equation}
The solution of~(\ref{eq:RGe}) yields the Coulomb phase which is
also obtained from the operator~$U_\Omega$ acting on the two particle
final state. In fact, the operator~$U_\Omega$ satisfies the
renormalization group equation~(\ref{eq:RGe})
\begin{equation}
\left( \mu \frac{\partial}{\partial \mu} + i G \right)
U_\Omega = 0 .
\end{equation}
with the imaginary part of the anomalous dimension replaced by
the operator~$G$
\begin{equation} \label{eq:RGeU}
G       =  - \frac{\alpha}{2} \int d^3 \vec{v} d^3 \vec{v}\,'
           \frac{vv'}{\sqrt{(vv')^2-1}} {:}n(v)n(v'){:}.
\end{equation}
Note that the hermiticity of~$G$ leads to a unitary operator~$U_\Omega$.
This makes manifest the connection of the Coulomb phases with the
imaginary parts of the anomalous dimensions. In particular, this
argument shows that it is appropriate to transfer the phases generated
by the imaginary parts of the anomalous dimensions from the
Wilson coefficients to the states by applying the
operator~$U_\Omega$ to the states.

Unfortunately, life is not that simple in the nonabelian case. The
currents live in the adjoint representation of the gauge group and are
given by
\begin{equation} \label{eq:jmua}
j_\mu^a (x) = \bar{Q}_\alpha (x) T^a_{\alpha\beta} \gamma_\mu Q_\beta (x),
\end{equation}
where~$T^a$ is the generator of color~$SU(3)$ in the fundamental
representation.
They become in the nonrecoil approximation
\begin{equation}
j_\mu^a (x) \to J_\mu^a (x) =  \int\! d^3 \vec{v}\, v_\mu\,
          \delta^3 (\vec{x} - \vec v x_0 / v_0) \,n^a(v),
\end{equation}
where now
\begin{equation}
n^a(v) =  \sum_{s=\pm} \left(
          b^\dagger_\alpha (v,s) T^a_{\alpha \beta} b_\beta (v,s)
        - d^\dagger_\alpha (v,s) T^{a*}_{\alpha \beta} d_\beta (v,s)
        \right).
\end{equation}
Due to the color structure these currents do not commute any more and
an exact solution is not possible. Nevertheless we shall proceed along
the lines suggested by the abelian case and define a nonabelian
counterpart of the operator~$G$ which is given to order~$\alpha_s$ by
\begin{equation}
{\cal G} = - \frac{\alpha_s}{2} \int d^3 \vec{v} d^3 \vec{v}\,'
           \frac{vv'}{\sqrt{(vv')^2-1}} {:}n^a (v)n^a (v'){:} .
\end{equation}
This reproduces the imaginary parts of the anomalous dimensions
discussed in the previous section
(cf.~(\ref{eq:imag})) when applied to a two particle state.
Furthermore,~${\cal G}$ is a hermitian operator and a
renormalization group equation analogous to~(\ref{eq:RGeU}) may be
used to define the nonabelian generalization~${\cal U}_\Omega$
of the operator~$U_\Omega$. However, the self-coupling of the gluons
and the existence of light quarks imply a nontrivial~$\beta$ function.
Thus the equation determining~${\cal U}_\Omega$ in the nonabelian
case is
\begin{equation} \label{eq:NARGe}
\left( \mu \frac{\partial}{\partial \mu} +
       \beta \frac{\partial}{\partial \alpha}
       + i {\cal G} \right)
{\cal U}_\Omega = 0 .
\end{equation}
As in the abelian case the operator~${\cal U}$ is a unitary operator
because~${\cal G}$ is hermitian.
The solution of~(\ref{eq:NARGe}) may be obtained in closed form and
is given by
\begin{equation}
{\cal U}_\Omega  =  \exp \left\{ - i \frac{\alpha_s(\mu^2)}{2}
                    \int d^3 \vec{v} d^3 \vec{v}\,'
                   \frac{vv'}{\sqrt{(vv')^2-1}} {:}n^a(v)n^a(v'){:}
                   \ln \left(\frac{m_Q}{\mu} \right)  \right\} .
\end{equation}
This operator again defines a unitary transformation on the
multiparticle states. Accordingly this nonabelian generalization
of the Coulomb phase may be shifted from the Wilson coefficients to
the states, rendering the anomalous dimensions real to order~$\alpha_s$.

Unlike the abelian case this is not the end of the story since there
are higher order contributions to the anomalous dimensions. In general,
the anomalous dimensions will contain imaginary parts in all orders
of~$\alpha_s$. Similar to the lowest order case one may define an
anomalous dimension operator by replacing the color factors by
appropriate density operators in the following way. The color factor
of some set of diagrams is given by a tensor product of~$T^a$'s
corresponding to heavy quark-gluon vertices. However in the nonrecoil
approximation every~$T^a$ may be assigned to a fixed velocity~$v$.
Thus an anomalous
dimension operator may be defined by replacing the tensor product of
the~$T^a$'s by a normal ordered product of the density operators~$n^a(v)$.
The generalization of~${\cal G}$ to all orders may be obtained by
picking out the antihermitian part of this operator, which exactly
corresponds to the imaginary part of the anomalous dimension. The
hermitian operator~${\cal G}$ is now obtained by multiplying the
antihermitian part of this anomalous dimension operator by~$(-i)$.

The kernel~${\cal G}$ is put into the renormalization group
equation~(\ref{eq:NARGe}) and the solution defines a unitary
operator~${\cal U}_\Omega$
which may be used to redefine the states, thereby
removing the imaginary parts of the anomalous dimension to any
desired order.

${\cal G}$ may be expanded in powers of~$n^a (v)$ corresponding to
two-, three-,~$\ldots$, $n$-particle operators
\begin{eqnarray}
{\cal G}  &=& \frac{1}{2!} \int d^3 \vec{v}_1 d^3 \vec{v}_2
              f_{(2)}^{ab} (v_1, v_2) {:}n^a(v_1)n^b(v_2){:}
\\ \nonumber
          && \mbox{} + \frac{1}{3!}
            \int d^3 \vec{v}_1 d^3 \vec{v}_2 d^3 \vec{v}_3
            f_{(3)}^{abc} (v_1, v_2, v_3)
           {:}n^a(v_1)n^b(v_2) n^c (v_3){:}
          + \cdots .
\end{eqnarray}
The kernels~$f_{(n)}^{a_1\cdots a_n}$
of these operators
are power series in~$\alpha_s$ with the property
\begin{equation}
  f_{(n)}^{a_1\cdots a_n} = {\cal O} (\alpha_s^{(n-1)}).
\end{equation}
Furthermore, the~$f_{(n)}^{a_1\cdots a_n}$
may be decomposed into rank~$n$ invariant~$SU(3)$
tensors; thus the lowest term~$f_{(2)}^{ab}$ must be proportional
to~$\delta^{ab}$
\begin{equation}
f_{(2)}^{ab} = - \delta^{ab} \alpha_s
               \frac{vv'}{\sqrt{(vv')^2-1}} + \cdots,
\end{equation}
where the dots indicate higher order terms.
Up to now only the leading and next-to-leading contributions
to~$f_{(2)}^{ab}$ have been calculated~\cite{GKMW91,KMM92}.

\section{Potentials and Phases}
\label{sec:pot}

Finally we shall point out the relation between the phase
operator~${\cal U}_\Omega$ and the nonrelativistic interquark
potential. In order to do this we consider a final state
consisting of a heavy quark and a
heavy antiquark in the singlet state with velocities~$v\approx v'$.
The generalization to other two heavy quark final states is obvious
since they differ only by color factors.

We shall consider the nonrelativistic limit, in
which~$v_0, v_0^\prime \approx 1$ and we have
\begin{equation}
  \frac{ v\cdot v'}{\sqrt{ (v\cdot v')^2 - 1}} \approx
  \frac{1}{|\vec u|} = \frac{1}{u},
\end{equation}
where~$\vec u$ is the
velocity of one quark in the rest frame of the other.

In this limit one obtains a simple result for the imaginary part
of the
anomalous dimension up to two loops for this final state
which reads~\cite{KMM92}
\begin{equation}
  \Imag \gamma(u) = \alpha_s C_F
                \left[ 1 + \frac{\alpha_s}{4\pi}\left( \frac{31}{9}C_A
                        - \frac{10}{9}n_f\right)\right]
                \frac{1}{u} + \cdots,
\end{equation}
where the dots represent terms which are finite as~$u \to 0$.
$C_F=4/3$ and~$C_A=3$ are the eigenvalues of the Casimir operator
in the fundamental and the adjoint representation respectively.
This imaginary part of the anomalous dimension will create a phase
of the Wilson coefficient which may be absorbed into the states
by applying the operator~${\cal U}_\Omega$ defined in the last section.

The connection between the potential and the phases appearing in the
Wilson coefficient proceeds along the lines well known from the QED
case~\cite{Dol64,KF70}. Here the Coulomb phases may be traced back to the
long range
part of the potential. In order to connect the phases with the potential
we consider a nonrelativistic system of two heavy particles interacting
by a spherically symmetric potential. The Hamiltonian is
\begin{equation}
  H = H_0 + H_I = \frac{\vec p_1^{\;2}}{2m_1} + \frac{\vec p_2^{\;2}}{2m_2}
                        + V(|\vec r_1 - \vec r_2|).
\end{equation}
Separating the cms motion, the time evolution for the relative motion
is given in the interaction picture by
\begin{equation}
  U(t,t_0) = \exp\left( -i\int_{t_0}^t d\tau\, V(r_{12}(\tau))\right),
\end{equation}
where~$r_{12}(t)$ is the operator for the relative coordinate, again
in the interaction picture.

Since the two particles are very heavy,
we can have the particles simultaneously in a state of definite velocity and
definite position. We choose
\begin{equation}
  r_{12}(t) = u t,
\end{equation}
where~$u$ is the velocity operator. When acting on states containing
particles with definite velocities, the time evolution becomes a pure phase
\begin{equation}\label{coulomb phase}
  U(t,t_0) = \exp\left( -i\int_{t_0}^t d\tau\, V(u\tau) \right).
\end{equation}
Introducing the Fourier transform of the potential
\begin{equation}
  V(r) = \frac{1}{2\pi^2 r}\int_0^\infty dq\, q\, \tilde V(q)\, \sin(q r)
\end{equation}
we can rewrite~(\ref{coulomb phase}) as
\begin{equation}
  U(t,t_0) = \exp\left( -\frac{i}{2\pi^2 u}\int_{t_0}^t d\tau\int_0^\infty dq\,
                \frac{q\,\tilde V(q)}{\tau}\sin(q u\tau)\right).
\end{equation}
For a potential falling off only like~$1/r_{12}$,
the~$\tau$ integration is divergent for~$t \to \infty$ as well
as for~$t_0\to 0$. However, as argued before,~$t_0$ has to be cut
off at times of the order of~$1/m_Q$, since the effective theory
is not valid for shorter times/larger scales. The other cut off is
related to the renormalization point~$\mu$ by~$t \sim 1/\mu$.
These cut offs may be shifted from the time integration to the
integration over the variable~$q$
\begin{equation}
  \int_{t_0}^t\!d\tau\,\int_0^\infty\!dq\,
                \frac{q\,\tilde V(q)}{\tau}\sin(q u\tau) \mapsto
  \int_{1/t=\mu}^{1/t_0=m_Q}\!dq\,\int_0^\infty\!d\tau\,
                \frac{q\,\tilde V(q)}{\tau}\sin(q u\tau)
\end{equation}
and the~$\tau$ integration may be
performed to yield
\begin{equation}
  U(\mu;m_Q) = \exp\left( -\frac{i}{4 \pi u}
                      \int_{\mu}^{m_Q} \frac{dq}{q} \,
                q^2\,\tilde V(q)\right).
\end{equation}

The exponent is a dimensionless quantity and hence~$q^2 \tilde V(q)$
is also dimensionless. Calculated from QCD it is thus
some function
\begin{equation}
  \phi (\alpha_s) = q^2 \tilde V(q).
\end{equation}
However,~$\alpha_s$ is scale dependent and this dependence will render
the potential nontrivial.
Differentiation with respect to the scale~$\mu$ yields the
renormalization group equation~(\ref{eq:NARGe}) with the anomalous
dimension being purely imaginary
\begin{equation}
\Imag \gamma (\alpha_s) = - \frac{1}{4 \pi u} \phi (\alpha_s).
\end{equation}
This equation finally relates the imaginary part of the anomalous
dimension to the long range interquark potential for heavy
nonrelativistic quarks.

Including the scale-dependence of~$\alpha_s$, the potential may then
be {\em defined\/} as
\begin{equation}
\label{eq:pfg}
\tilde V (q) = \frac{\phi (\alpha_s (q^2))}{q^2}
             = - \lim_{u\to0}
               \frac{4 \pi u \Imag \gamma (\alpha_s (q^2))}{q^2}
\end{equation}
and we obtain from perturbation theory up to second order in the
particle-antiparticle singlet channel
\begin{equation}\label{potential}
  \tilde V(q) = -\frac{4\pi\alpha_s(q^2)\,C_F}{q^2}
        \left[ 1 + \frac{\alpha_s(q^2)}{4\pi}
                \left( \frac{31}{9}C_A - \frac{10}{9}n_f\right)\right],
\end{equation}
where the two loop running coupling constant has to be used. The
QCD potential between two static color sources has been
calculated long ago~\cite{Fis77,BT81}. Our expression
for the potential~(\ref{potential}), valid in the~$\overline{\rm MS}$
scheme, is in complete agreement with these
calculations.

Finally we point out that for dimensional reasons
nontrivial behavior of the potential
can only be generated by the scale dependence of the strong coupling
constant. For instance, the linear confinement
potential~$V_{{\rm lin.}}(r) \propto r$, which may be obtained from
lattice calculations and quarkonia-spectroscopy, corresponds to
nonperturbative~$\beta$-functions.  The Fourier-transform of the
linear potential is
\begin{equation}
\label{eq:Vlin}
    \tilde V_{{\rm lin.}} (q) = C \frac{1}{q^2}
     \frac{\Lambda^2}{q^2},
\end{equation}
where~$C$ is some numerical constant.
At this point an additional scale~$\Lambda^2$ has to enter the game,
which has to be generated by dimensional transmutation
{}from the scale dependence of the coupling constant.  In fact, the
general renormalization group equation
\begin{equation}
  \mu^2 \frac{\partial}{\partial\mu^2} \alpha(\mu^2)
    = \beta(\alpha(\mu^2))
\end{equation}
has the solution
\begin{equation}
  \frac{\mu^2}{\Lambda^2} = \exp \left(
    \int^{\alpha(\mu^2)}_{\alpha(\Lambda^2)}\!
    \frac{d\alpha}{\beta(\alpha)} \right).
\end{equation}
The potential~(\ref{eq:Vlin}) can now be expressed in terms of
the~$\beta$-function
\begin{equation}
 \tilde V_{{\rm lin.}} (q) = C \frac{1}{q^2} \exp \left(
    - \int^{\alpha(\mu^2)}_{\alpha(\Lambda^2)}\!
    \frac{d\alpha}{\beta(\alpha)} \right)
\end{equation}
and we can use equation~(\ref{eq:pfg}) backwards to deduce the
imaginary part of the anomalous dimension corresponding to the linear
potential
\begin{equation}
  \Imag \gamma_{{\rm lin.}} = - \frac{C}{4\pi^2u} \exp \left(
    - \int^{\alpha(\mu^2)}_{\alpha(\Lambda^2)}\!
    \frac{d\alpha}{\beta(\alpha)} \right).
\end{equation}
In principle, we could now reconstruct the real part of the anomalous
dimension from a dispersion relation.  However, since the
nonrelativistic potential approach breaks down for large~$vv'$, we
do not have any information in this region.

\section{Conclusions}
\label{sec:concl}

The imaginary parts of the anomalous dimensions of HqEFT may be
removed by a suitable redefinition of the multiparticle states.  The
redefinition of the multiparticle states is achieved by applying a
unitary operator, similar to the Coulomb-phase operator in QED.
While this phase operator can be constructed exactly in the QED case,
we can only give a perturbative construction of the nonabelian
analogue of the Coulomb-phase operator.  Since the phase operator
removes the imaginary parts of the anomalous dimensions, it depends on
the renormalization scale; the dependence is governed by a
renormalization group equation with the imaginary parts of the
anomalous dimensions only.

Furthermore we have elucidated the connection between the imaginary
parts of anomalous dimensions and the long range inter-quark
potential.  The potential derived from the HqEFT calculation in the
two-particle sector up to two loops coincides with earlier
perturbative calculations.

We believe that we have settled the problem of the apparently complex
anomalous dimensions in the multiparticle sector of HqEFT.  The
imaginary parts are an artifact of an improper definition of the
states; after a suitable redefinition of the states there is no such
thing as a complex anomalous dimension.

\section*{Acknowledgements}

We thank Howard Georgi, Matt McIrvin, and Panagiotis Manakos for
enlightening discussions.
This research was supported in part by the National Science Foundation
under Grant \#PHY-8714654 and by the Texas National Research Laboratory
Commission under grant RGFY9206.
W.~K.~acknowledges financial support by Studienstiftung des Deutschen
Volkes.
T.~O.~acknowledges financial support by
Deutsche Forschungsgemeinschaft under Grant \#~Oh~56/1-1.


\end{document}